# The Algorithm Steering and Trigger Decision mechanism of the ATLAS High Level Trigger

G. Comune[(*)], A. Corso-Radu
*LHEP, Bern, Switzerland*

M. Elsing, M. Grothe, T. Schoerner-Sadenius
*European Laboratory for Particle Physics, Geneva, Switzerland*

D. Wicke

*Fachbereich Physik, Gesamthochschule Wuppertal, Bergische Universitaet, Germany*

S. George, A. Lowe
*Department of Physics, RHBNC, University of London, Egham, United Kingdom*

T. Shears
*Department of Physics, Liverpool University, United Kingdom*

J. T. Baines
*Rutherford Appleton Laboratory, Chilton, Didcot, United Kingdom*

S. Gonzalez
*Department of Physics, University of Wisconsin, Madison, Wisconsin*

on behalf of the ATLAS HLT group[1]

An algorithm Steering and Data Navigation mechanism for the High Level Trigger Selection Software in the ATLAS experiment at the LHC proton-proton collider is presented. Relevance is given to an event selection strategy where the reconstruction is limited to detector geometrical Regions of Interest where high momentum signals are detected. The proposed reconstruction process proceeds in incremental steps and event rejection can take place at any of these steps.

## 1. INTRODUCTION:

ATLAS is one of the four experiments under construction at the Large Hadron Collider (LHC) facility at the European Organization for Nuclear Research (CERN) in Geneva. ATLAS is a general-purpose detector designed to have nearly 4π geometry around the interaction point.

LHC is designed to collide protons (and heavy ions) at a Center of Mass of 14 TeV with a bunch cross rate of 40 MHz with an average event size of 1.6 MB.

At the very high design luminosity of $10^{34}$ cm$^{-2}$s$^{-1}$ the p-p interaction rate is about 1GHz. Given the LHC design characteristics and the wide physics potential of the ATLAS detector [12] it is clear that a very effective event selection strategy is needed.

A three level Trigger system is under investigation and development (Figure 1):

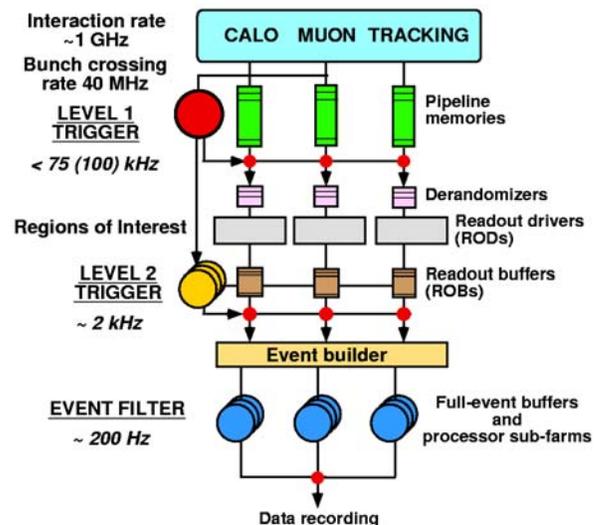

Figure 1: The ATLAS Trigger.

---

[(*)] Presenter at the conference

-The Trigger Level-1 (LVL1) is a hardware system based on dedicated electronics and pipelined memories. It is capable of reducing the initial 40 MHz interaction rate down to 75 kHz based on high $P_T$ signals coming from the Calorimeters and Muon subsystems

-The Trigger Level-2 (LVL2) is based on a farm of commodity processor nodes running software algorithms developed to achieve event rejection/selection with an average latency of 10ms.

Access to the event data is going to be guided by the geometrical information provided by the LVL1 in terms of Regions of Interest (RoIs) where the first Trigger level has already found interesting physic signal. By limiting the reconstruction to the RoIs, when possible, the access to the event data is reduced on average to few per-cent of the full event dramatically reducing the networking and computing power needed.

The LVL2 reduces the rate from the incoming 75 kHz, feed by the LVL1, to ~2 kHz.

-The Trigger Level-3 or Event Filter (EF) is built on a farm of commercial processor nodes running event selection software based on the offline reconstruction.
The EF has access to the full event and the reconstruction is guided by the selection result of the LVL2. The EF reduces the rate from ~2 kHz to the final ~200 Hz with an average latency of ~1s. The selected events are then recorded on mass storage for later offline analysis. EF together with LVL2 forms what is known as High Level Trigger (HLT).

The RoIs found in the LVL1 will *steer* the execution of the appropriate algorithms run in the LVL2. Different types of RoIs are found at the LVL1 (Electromagnetic, Muon, Hadronic…) so different sets of algorithms will be executed in the LVL2 according to the RoI type. At the end of LVL2 selection if an event is accepted a LVL2 result is produced and used to guide ("*steer")* the EF algorithm execution.

In this paper we will introduce a key concept, central to the ATLAS HLT: a stepwise algorithm Steering and decision mechanism based on a data navigation schema that fully implements the concept of RoI guided reconstruction.
The purpose of an algorithmic execution in steps is to achieve an early rejection of any unwanted event as soon as some physics conditions are not meet at any step of the decisional process.
A data navigation mechanism is proposed with the intent to allow any algorithm to access reconstructed objects produced (*seeds*) at any previous step without the need for the algorithm to be aware of the details on how those objects were created. The initial *seeds* for the LVL2 step processing are the LVL1 RoIs while for the EF the LVL2 Result contains the initial *seeds*.

## 2. THE ATLAS HIGH LEVEL TRIGGER SELECTION SOFWARE

### 2.1. HLT overview

As already mentioned the LVL2 and EF Trigger levels are implemented on commodity processor nodes running a commercially available operating system.
The LVL1 searches for RoIs and produces a result containing information about all primary and secondary RoIs found in a particular event.
LVL1 Result is sent to one of the LVL2 processor nodes through the ATLAS T/DAQ [2] infrastructure.

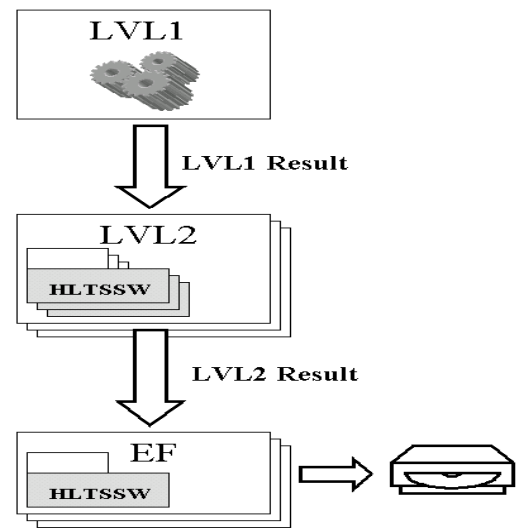

Figure 2: The Trigger Data Flow

At the HLT event selection resumes "*seeded by*" the RoIs found by the LVL1.
The LVL2 processing nodes have in principle access to the full event data but due to the very demanding time constraints data access is limited, where possible, to the detector geometrical regions described by the RoI. By doing so the amount of data that needs to be transferred over the network and analyzed remains very limited.
The LVL2 Trigger selection takes place and if any physics signatures are satisfied by the event, it is accepted and sent to the EF with the LVL2 Result appended.
The LVL2 Result contains the criteria that have been satisfied by the event to be selected and the relevant objects that are going to be used to guide the EF selection.
In a fashion very similar to the LVL2 the EF resumes the event selection "*seeded by*" the LVL2 information contained in the LVL2 Result and it will execute only the algorithms that are needed by the signatures that caused the event accept at LVL2 i.e. if at LVL2 the event is found to contain only electrons no muon identification algorithms will be executed; it will be shown later how this is achieved.

The EF has access to the full event data and can, and probably will, access the full event data to achieve event selection.

The software responsible for the event selection is called the HLT Selection Software (HLTSSW)

The current HLTSSW implementation draws entirely on Athena [5,6], the ATLAS offline framework.

Central to Athena is the transient event data store, StoreGate [13], which realizes the "blackboard" architecture of the framework.

Objects are recorded in StoreGate with a key by a client and can be retrieved by key from any other client algorithm later in the reconstruction/selection chain.

Before an event is processed, seed objects (LVL1 RoIs for LVL2 or LVL2 Result for EF) are created in the transient store and used to guide the selection process.

Externalization of data and objects is achieved through a Byte Stream (BS) converter technology

An object that needs to be externalized and sent to the next Trigger level or to mass storage is converted to its byte stream representation and appended to the event to be shipped with it.

A specific conversion algorithm has to be implemented for the object to be externalized. The algorithm has to create a serialized object representation. Byte Stream technology takes care of appending the serialized object representation to the event data and is responsible for the handling of the persistency technology.

## 2.2. The HLTSSW

HLTSSW is responsible for accepting events from the previous Trigger level, for the result unpacking and for the event selection or rejection.

HLT nodes run the HLTSSW. Given the LVL2 latency (10ms) the normal operating system time slice is totally inadequate for the performance demands imposed by the LVL2 so a multithreaded execution environment has been put in place. This imposes some restrictions on the algorithm development and in general on the porting of offline like software into an online environment [7].

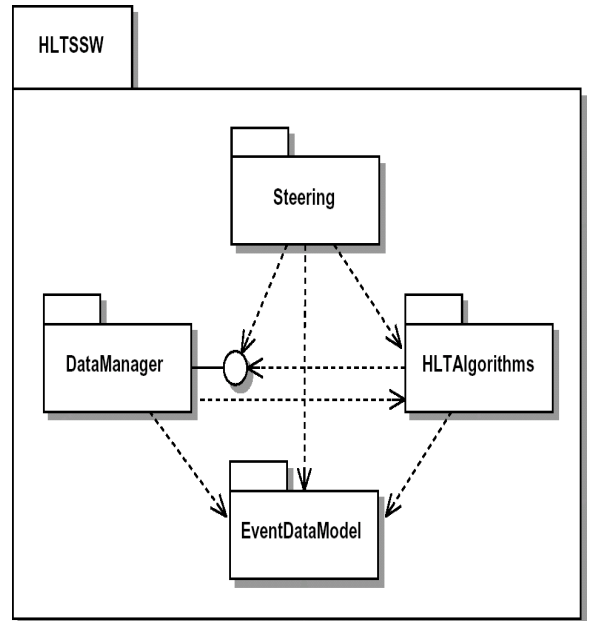

Figure 3: The HLTSSW

The package responsible for the event selection is called Steering and it is in Steering that the stepwise algorithmic execution takes place.

The HLT Configuration fixes the set of algorithms and conditions that are to be executed and checked by the Steering in a particular run. The configuration is dynamic and takes place at run start.

The decisional process is broken up into incremental steps; the rejection can occur at every step of the chain.

For an event to be accepted, proper conditions have to be satisfied at all the steps.

## 3. THE CONFIGURATION

At the beginning of every run the HLTSSW is initialized and configured based on the output of a configuration stage [14].

The configuration stage returns to the HLTSSW specific information regarding what Algorithms (Sequences) are to be executed and what conditions (Signatures) are to be checked.

A Signature corresponds to a set of physics criteria sufficient to trigger the event while a Sequence corresponds to the set of algorithms have to be executed to check if those criteria apply.
The configuration is based on XML files

## 3.1. Trigger Elements

A Trigger Element represents a "hypothesis" that has to be confirmed or rejected.

Signatures and Sequences are built upon Trigger Elements (TEs) (Figure 4).

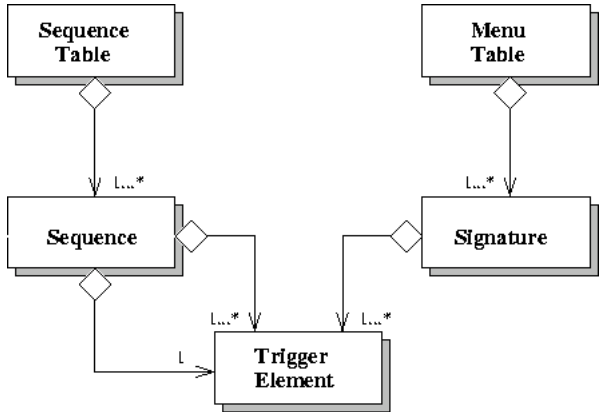

Figure 4: The Configuration Classes

A string uniquely identifies the Trigger Element (Figure 5). This string is in general chosen to represent the physical feature associated with that TE (*hypothesis*). TEs are created by the Steering based on configuration information and can be "Active" or "Inactive". This property is the real important information used by the Steering to take appropriate decisions during the event processing. Trigger Elements are "*hypothesis*" that have to be confirmed or discarded by algorithms. The hypothesis represented by a TE is signaled as satisfied to the Steering through the "Active" flag.

```
TriggerElement
-m_active bool
-m_label string

+TriggerElement(string)
+label() : string
+isActive() : bool
+setActive(bool) : void
```

Figure 5: The Trigger Element

## 3.2. Signatures

Signatures are basically lists of Trigger Elements (Figure 4). Signatures represent particular physical criteria that need to be satisfied in order for an event to be accepted. An example of a possible Signatures is:

*"e50i" + "e50i"*

which indicates "two electrons with 50GeV with isolation criteria applied". This Signature is satisfied if two TEs with label "*e50i*" are found active in the event store. Signatures are the entities used by the Steering to achieve event selection. At every step the Steering checks Signatures and the dynamic transient event store counting the number of existing active TEs with the required "label"

## 3.3. Sequences

Sequences carry the information needed by the Steering in order to perform the algorithm sequencing. An example of a possible Sequence is:

*"e50"*     *"EMIsolation"*     *"e50i"*

the second string being the name of the isolation criteria checking algorithm that is to be executed if a TE with label "e50" is found active in the event store. The third string is the label of the TE that is created by the Steering before executing the algorithm. This TE represent the isolation *hypothesis* and will signal the successful "*hypothesis*" if set to "Active" by the algorithm.

## 3.4. Menu and Sequence Tables

Signatures and Sequences are grouped together in Tables to be used by the Steering.

Configuration creates those tables and passes the full list of tables at initialization time to the Steering as a collection of pairs of Signature and Sequence tables.

There is a pair of Sequence and Signature (Menu) Tables for every step.

The Steering at a given step uses one Sequence and one Menu Table.

## 4. NAVIGATION

The guiding principle in the Trigger event selection is the RoI based reconstruction.

The LVL1 searches the event for high $P_T$ signals in the Calorimeters and in the Muon spectrometer. If any of these signals is found a RoI is built and appended to the LVL1 Result. The result is externalized and sent to a LVL2 processing node together with the event data.

The Steering takes ownership of the event and unpacks the LVL1 Result recreating the RoIs in the event store. For every RoI a corresponding Trigger Element is created and recorded in StoreGate and a "*uses*" data navigation link is set up between the two (Figure 6a).

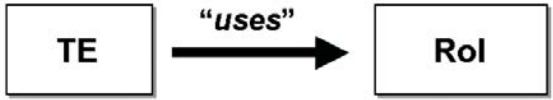

Figure 6a: Navigation link

There are different RoI *types* (Electromagnetic, Muon, Hadron…) so different TE types will be created to reflect this fundamental difference between RoIs.

After this initial setup is concluded the stepwise processing takes place and based on the type of TE (RoIs) found in the event store the appropriate set of algorithms is invoked based on the Configuration information. At the first step algorithms will *navigate* back from the hypothesis TE to the "seed" TE and retrieve the RoI (Figure 6b).

Once the algorithm has retrieved the RoI it will use the geometrical information coming with it, and any other information encoded by the LVL1 into the RoI, to refine and eventually accept the LVL1 signal.

If the algorithm strategy allows it, reconstruction takes place in the geometrical region identified by the RoI limiting the amount of data preparation, transfer and processing on average to few percent of the full event.

It is clear now that the same algorithm has to be able to (and will) run several times per event because of the RoI guided reconstruction approach.

Downstream algorithms will always be able to navigate back to the initial *seeding* RoI making extensive use of the RoI approach at any step of the selection chain.

The RoI is not the only object that can be used by the algorithms for the seeding mechanisms but any other reconstructed object produced by any previous algorithm can be used, extending the *seeding* concept to any object that can, for the information it carries, act as a *seed* for another algorithm.

The details on how the objects (*seeds*) have been created are irrelevant to the client algorithms because they can always navigate back to the seed (Figure 6b) and retrieve the objects that are needed, this is the basic idea of a "blackboard" event store.

Logical relations like "*seeded by*", "*uses*" or "*excludes*" can link TEs and objects.

In the current implementation the "*seeded by*" relation is set up by the Steering and embodies the *seeded* reconstruction idea. Other relations, like *uses*, can be set up and used by algorithm(s).

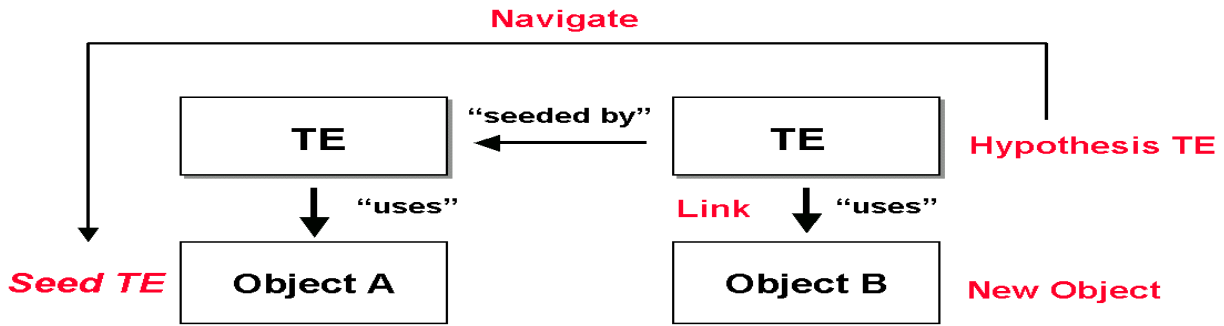

Figure 6b: The data navigation

### 4.1. History Objects

All the navigational relations between TEs and Objects are encoded in objects called History Objects.

History Objects are STL hash multimaps that map a string (the "relation") and an instance of a templated Holder of the object's pointer.

The string relation is completely arbitrary and the meaning or use of it is handled and set by the Steering or by the algorithmic code.

Figure 7 shows the actual content of the History Object in the case of a TE linked to three different objects with three different relations.

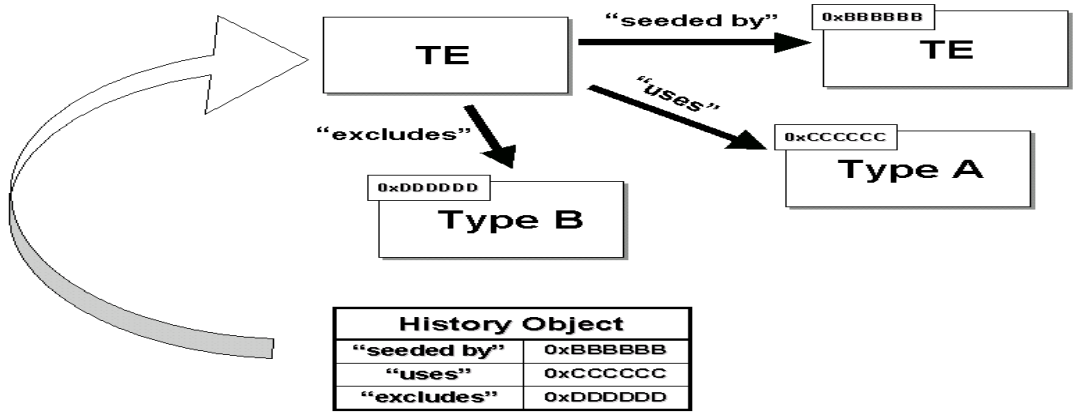

Figure 7: The History Object

## 5. ALGORITHMS

Algorithms are in charge of activating the hypothesis TE that they receive in input from the Steering.

The flow of actions relevant to the Steering that takes place during the algorithm execution is shown in Figure 8.

In the same picture are shown the actions related to the data navigation.

An algorithm receives the hypothesis TE and from there it navigates back to the *seed* TE (Figure 6b), it retrieves the seed (RoI or any other object) and after having created some reconstructed objects it sets up the "*uses*" relation between the input TE and the objects.

Before returning control to the Sequencing algorithm in the Steering it activates the *hypothesis* TE signaling the satisfied condition.

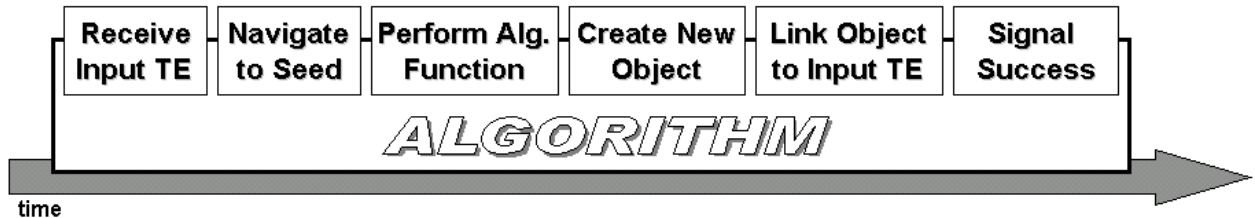

Figure 8: Algorithm execution

## 6. THE STEERING

The Steering is one of building block of the HLTSSW and together with the Data Manager is known as the "core software" [10,11].

Figure 9 shows how the stepwise reconstruction works for all the steps in a Trigger level.

At the LVL2 the TEs that pass the selection, together with the linked objects, are externalized and included in the LVL2 Result that is sent to one of the EF farm nodes. The same figure applies almost unmodified to LVL2 and to EF with the prescription that for the EF the initial *seeds* are the TEs relevant for the Signatures that passed the LVL2 selection.

The objects linked to those initial TEs are not the LVL1 RoIs but the reconstructed objects that where linked to the "successful" TEs at the last LVL2 step before the event was accepted.

### 6.1. The Step Decision

The decision that occurs at every step in Figure 9 is taken based on the number of active TE existing in the event store.

The Step Decision algorithm of the Steering counts the number of active TEs and if they are enough to satisfy at least one of the Signatures in the step Menu Table then the event is accepted to the next selection step.

Events that pass all steps in a Trigger level pass the selection and are either sent to the next level or to mass storage for offline analysis. A fraction of the events is accepted forcedly (forced accept) to provide offline analysis with an unbiased sample of events. If for some unforeseen reason one of the Signatures gives place to an unwanted event accept rate only a percentage of the events is Preselected and really accepted.

The amount of Forced Accept and Preselected events is configurable via the HLT Configuration.

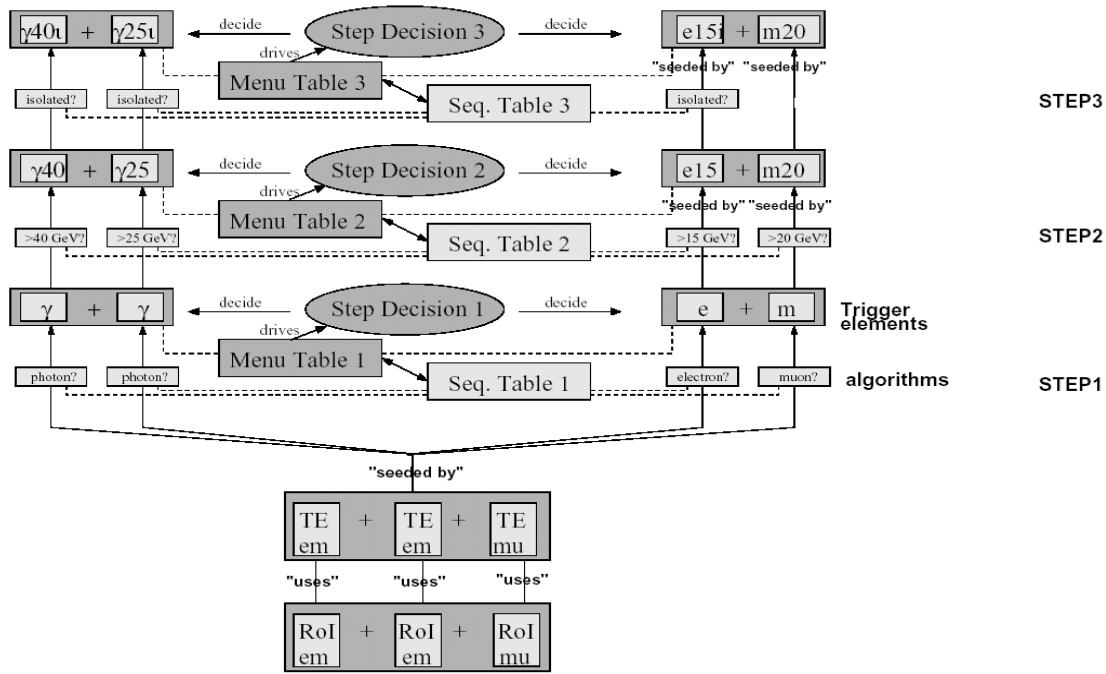

Figure 9: the stepwise selection process

## 7. CONCLUSIONS

An approach to event selection in a very demanding environment like ATLAS as been described. The *core* selection software has to address two major challenges imposed by the LHC environment: performance and complexity.

The *seeded* reconstruction addresses the performance aspect allowing algorithms to focus only on the promising event regions or objects already pinpointed at any previous Trigger level or step. Stepwise reconstruction allows early rejection of unwanted events as early as they show not to fulfill any of the Signatures in the Menu Table saving processing time for those events that on the contrary fulfill such criteria.

The Framework has shown not to introduce any major overhead to the selection process while addressing the demand in flexibility imposed by the complex nature of a hadronic collider and the broad range of Physics that will be under investigation at ATLAS.

## References

[1] The ATLAS High Level Trigger group[a] http://atlas.web.cern.ch/Atlas/GROUPS/DAQTRIG/HLT/AUTHORLISTS/chep2003.pdf

[2] HP. Beck et al., ATLAS TDAQ, a network-based architecture, TDAQ Data Collection note 59

[3] C. Bee et al., Event Handler Requirements, https://edms.cern.ch/document/361786/1

[4] C. Bee et al., Event Handler Design, https://edms.cern.ch/document/367089/1.1

[5] ATHENA manual, http://atlas.web.cern.ch/Atlas/GROUPS/SOFTWARE/OO/architecture/General/Tech.Doc/Manual/2.0.0-DRAFT/AthenaDeveloperGuide.pdf

[6] Gaudi manual, http://proj-gaudi.web.cern.ch/proj-gaudi/GDG/v2/Output/GDG.htm

[7] Guidelines for writing thread-safe LVL2 algorithms http://sarmstro.home.cern.ch/sarmstro/algorithms/guidelines/coding-for-LVL2.txt

[8] S. Gonzalez et al., Use of Gaudi in the LVL2 Trigger: The Steering Controller, https://edms.cern.ch/file/371778/1/daq-2002-012.pdf

[9] A. Bogaerts et al., https://edms.cern.ch/document/375305/1

[10] S. George et al., PESA high level trigger selection software requirements, ATL-DAQ-2001- 005

[11] M. Elsing et al., Analysis and Conceptual Design of the HLT Selection Software, ATL-DAQ-2002-01

[12] ATLAS TDR 15, CERN/LHCC 99-15

[13] P. Calafiura et al., The StoreGate: a Data Model for the Atlas Software Architecture [CHEP 2003 proceedings

[14] M. Elsing et al., Configuration of event-selection criteria in the ATLAS trigger system [CHEP 2003 proceedings]

[a]S. Armstrong, J.T. Baines, C.P. Bee, M. Biglietti, A. Bogaerts, V. Boisvert, M. Bosman, S. Brandt, B. Caron, P. Casado, G. Cataldi, D. Cavalli, M. Cervetto, G. Comune, A. Corso-Radu, A. Di Mattia, M. Diaz Gomez, A. dos Anjos, J. Drohan, N. Ellis, M. Elsing, B. Epp, F. Etienne, S. Falciano, A. Farilla S. George, V. Ghete, S. González, M. Grothe, A. Kaczmarska, K. Karr, A. Khomich, N. Konstantinidis, W. Krasny, W. Li, A. Lowe, L. Luminari, H. Ma, C. Meessen, A.G. Mello, G. Merino, P. Morettini, E. Moyse, A. Nairz, A. Negri, N. Nikitin, A. Nisati, C. Padilla, F. Parodi, V. Perez-Reale, J.L. Pinfold, P. Pinto, G. Polesello, Z. Qian, S. Rajagopalan, S. Resconi, S. Rosati, D.A. Scannicchio, C. Schiavi, T. Schoerner-Sadenius, E. Segura, T. Shears, S. Sivoklokov, M. Smizanska, R. Soluk, C. Stanescu, S. Tapprogge, F. Touchard, V. Vercesi, A. Watson, T. Wengler, P. Werner, S. Wheeler, F.J. Wickens, W. Wiedenmann, M. Wielers, H. Zobernig